\documentclass[twocolumn]{aastex631}

\newcommand{\gwsn}{\texttt{GWSkyNet}}
\newcommand{\gwsnm}{\texttt{GWSkyNet-Multi}}

\begin{document}

\title{Explaining the \gwsnm ~machine learning classifier predictions for gravitational-wave events}

\correspondingauthor{Nayyer Raza}
\email{nayyer.raza@mail.mcgill.ca}

\author[0000-0002-8549-9124]{Nayyer Raza}
\affiliation{Department of Physics, McGill University, 
3600 rue University, Montr\'{e}al, QC H3A2T8, Canada}
\affiliation{Trottier Space Institute at McGill, 
3550 rue University, Montr\'{e}al, QC H3A2A7, Canada}

\author[0009-0009-6826-4559]{Man Leong Chan}
\affiliation{Department of Physics and Astronomy, University of British Columbia, 
Vancouver, BC V6T1Z1, Canada}

\author[0000-0001-6803-2138]{Daryl Haggard}
\affiliation{Department of Physics, McGill University, 
3600 rue University, Montr\'{e}al, QC H3A2T8, Canada}
\affiliation{Trottier Space Institute at McGill, 
3550 rue University, Montr\'{e}al, QC H3A2A7, Canada}

\author[0000-0003-2242-0244]{Ashish Mahabal}
\affiliation{Division of Physics, Mathematics and Astronomy, California Institute of Technology, Pasadena, CA 91125, USA}
\affiliation{Center for Data Driven Discovery, California Institute of Technology, Pasadena, CA 91125, USA}

\author[0000-0003-0316-1355]{Jess McIver}
\affiliation{Department of Physics and Astronomy, University of British Columbia, Vancouver, BC V6T1Z1, Canada}

\author[0000-0001-5002-0868]{Thomas C. Abbott}
\affiliation{Department of Physics, McGill University, 
3600 rue University, Montr\'{e}al, QC H3A2T8, Canada}
\affiliation{Trottier Space Institute at McGill, 
3550 rue University, Montr\'{e}al, QC H3A2A7, Canada}

\author[0000-0003-2205-2912]{Eitan Buffaz}
\affiliation{Department of Physics, McGill University, 
3600 rue University, Montr\'{e}al, QC H3A2T8, Canada}
\affiliation{Trottier Space Institute at McGill, 
3550 rue University, Montr\'{e}al, QC H3A2A7, Canada}

\author[0000-0001-7815-7604]{Nicholas Vieira}
\affiliation{Department of Physics, McGill University, 
3600 rue University, Montr\'{e}al, QC H3A2T8, Canada}
\affiliation{Trottier Space Institute at McGill, 
3550 rue University, Montr\'{e}al, QC H3A2A7, Canada}

\shorttitle{Explaining \gwsnm ~predictions}
\shortauthors{Raza et al.}
 
\begin{abstract}

\gwsnm ~is a machine learning model developed for classification of candidate gravitational-wave events detected by the LIGO and Virgo observatories. The model uses limited information released in the low-latency Open Public Alerts to produce prediction scores indicating whether an event is a merger of two black holes, a merger involving a neutron star, or a non-astrophysical glitch. This facilitates time sensitive decisions about whether to perform electromagnetic follow-up of candidate events during LIGO-Virgo-KAGRA (LVK) observing runs. However, it is not well understood how the model is leveraging the limited information available to make its predictions. As a deep learning neural network, the inner workings of the model can be difficult to interpret, impacting our trust in its validity and robustness. We tackle this issue by systematically perturbing the model and its inputs to explain what underlying features and correlations it has learned for distinguishing the sources. We show that the localization area of the 2D sky maps and the computed coherence versus incoherence Bayes factors are used as strong predictors for distinguishing between real events and glitches. The estimated distance to the source is further used to discriminate between binary black hole mergers and mergers involving neutron stars. We leverage these findings to show that events misclassified by \gwsnm ~in LVK's third observing run have distinct sky area, coherence factor, and distance values that influence the predictions and explain these misclassifications. The results help identify the model's limitations and inform potential avenues for further optimization.

\end{abstract}


\section{Introduction} \label{sec:intro}

The Advanced LIGO \citep{Aasi2015} and Advanced Virgo \citep{Acernese2015} gravitational-wave detector network has discovered 90 significant events of merging compact binaries across their first three observing runs \citep[O1-O3,][]{Abbott2021_gwtc3}. These include mergers of two black holes (BBH), two neutron stars (BNS), and a neutron star and a black hole (NSBH). Since the start of the third observing run (O3), the LIGO-Virgo-KAGRA (LVK) collaboration \citep[KAGRA;][]{Akutsu2021} has also produced low-latency Open Public Alerts (OPAs) for significant gravitational-wave transient candidate events\footnote{\url{https://emfollow.docs.ligo.org/userguide/}}. These alerts are used by astronomers to perform follow-up searches for electromagnetic and neutrino counterparts, and to enable multi-messenger studies of compact binary merger events.

Since telescope time and resources for follow-up studies are limited, astronomers must make rapid decisions following an OPA on whether the candidate merits expenditure of these resources. This task is further complicated by the fact that not all candidate events identified by low-latency pipelines are actual merger events; a large number are transient instrumental or environmental noise in the detectors known as ``glitches''. Indeed, in O3 there were 77 OPAs issued for candidate compact binary coalescence (CBC) events, but more detailed analysis by the LVK collaboration revealed that only 43 (56\%) of these can be confidently classified as astrophysical in origin \citep[probability $p_{\mathrm{astro}} > 0.5$ in][]{Abbott2021_gwtc3}. Candidate events with $p_{\mathrm{astro}} < 0.5$ are more likely glitches, but could also arise from random Gaussian noise fluctuations, or even be a weak real signal.

As a complementary tool for helping astronomers make rapid follow-up decisions with limited data, \cite{Cabero2020} developed the \gwsn ~machine learning binary classifier for gravitational-wave events. This real-versus-noise classifier predicts whether a candidate gravitational-wave event is a real merger of compact objects or a non-astrophysical glitch, using only publicly available data released in OPAs. The \gwsn ~model as published in \cite{Cabero2020} has since been significantly refined and updated to produce even more robust predictions (Chan et al., in prep).

 Furthermore, even for real compact binary merger events, not all are expected to produce an electromagnetic (EM) counterpart. Mergers involving two neutron stars are the most promising, and are known to produce short gamma-ray bursts and kilonovae, as was observed with GW170817 \citep{Abbott2017_gw170817, Abbott2017_multimessenger}. The merger of a neutron star and black hole may produce an electromagnetic signal depending on the mass ratio of the two objects, but to-date no EM counterpart has been detected \citep[e.g.,][]{Vieira2020}. The merger of two black holes in the stellar mass range is not normally expected to produce any counterpart \citep[see, e.g.,][]{Branchesi2021}.
 
 \gwsnm ~\citep{Abbott2022} is an expansion of the \gwsn ~machine learning framework, and introduces a series of three one-versus-all classifiers that further distinguish sources as binary black hole (BBH) mergers, mergers involving at least one neutron star (NS), or glitches. The models use sky map information and associated metadata generated by the rapid localization pipeline \texttt{BAYESTAR} \citep{Singer2016}. With these inputs, \gwsnm ~performs one-vs-all classifications for the three source classes and produces three corresponding prediction scores as the probability for belonging to each class. Individually, these models have been shown to have test set accuracies of 94\% for the BBH-vs-all model, 94\% for NS-vs-all, and 95\% for glitch-vs-all. An overall accuracy of 93\% is achieved when using a multi-class hierarchical classification scheme as outlined in \cite{Abbott2022}. When \gwsnm ~is used to predict and classify the LVK OPAs released in O3, it correctly identifies 64/77 (83\%) of them in \cite{Abbott2022} (see Section \ref{subsec:O3_misclass} for a detailed discussion on updated predictions).

The fact that all three one-vs-all classifiers reach such high classification accuracies while utilizing only the limited information available in the OPAs points to the power and utility of deep learning techniques. Yet the trade-off for building complex models like the convolutional neural networks in \gwsnm ~is that they are not easily interpretable --- the many layers of the model make it difficult to understand what exactly the model is learning, how it is generating features from the input, and manipulating the input features to produce accurate outputs. This impacts our trust in the model and its predictions, because we cannot easily tell whether they are based on physically grounded truths, as opposed to artifacts that may have been introduced in the training process.

As machine learning techniques are increasingly employed by the astronomy community to facilitate fundamental discovery \citep[see for example the recent review by][]{Djorgovski2023}, it becomes important to understand complex ``black-box'' models and to establish whether they are consistent with our physical expectations of the systems they are modeling. Identifying the features that the models are focusing on can also drive new astrophysical understanding of the system \citep[e.g.,][]{Ntampaka2022}, improve the robustness of deep learning models used in data analysis \citep[e.g.,][]{Jadhav2023}, and provide clues to how the model architecture can be modified to abstract the underlying system more accurately \citep[e.g.,][]{Safarzadeh2022}.

In this paper we study the predictions of the \gwsnm ~machine learning model and identify the features that are most important in successfully differentiating between gravitational-wave source classes. In Section \ref{sec:methods} we describe the inputs to \gwsnm, how they are distributed, and our approach to studying them. In Section \ref{sec:results} we analyze the effects of these inputs on the predictions and how perturbing them changes the output. In Section \ref{sec:discussion} we discuss the implications of these results, providing insights into the underlying model, and using these insights to explain the misclassified events in O3. In Section \ref{sec:conclusion} we conclude.

\section{Methods} \label{sec:methods}

\subsection{The \textsc{\gwsnm} model}
The three one-vs-all models in \gwsnm ~take sky map information and associated metadata from the OPAs as generated by \texttt{BAYESTAR}, which runs once a candidate CBC event is identified by a low-latency detection pipeline. \texttt{BAYESTAR} uses the matched-filter detection pipeline output to compute Bayesian posterior probability distributions of the source’s sky position and distance to rapidly construct a 3D localization map. In particular it coherently models the signal response in the detector network, focusing on the amplitude, phase, and time delay on arrival of the signal at each detector \citep{Singer2016}. These also form the basis for computing the Bayesian evidences for the signal, noise, coherence, and incoherence models, and when compared give the evidence ratios, or Bayes factors, signifying support for one model compared to another.

The \texttt{BAYESTAR} data outputs used by \gwsnm ~are released in the form of a Flexible Image Transport System (FITS) file, which includes:
\begin{enumerate}
    \item a 2D sky localization probability map,
    \item a 3D volume probability projected onto 3 orthogonal plane images, 
    \item the mean and maximum distance to the source (with the maximum defined as mean + 2.5$\sigma$),
    \item the logarithm of the signal to noise Bayes factor (Log BSN),
    \item the logarithm of the coherence to incoherence Bayes factor (Log BCI), and
    \item the detector network involved in the detection.
\end{enumerate}
As detailed in \cite{Abbott2022}, these data are processed, projected and normalized before they are ready to be used as inputs to \gwsnm.

In \cite{Abbott2022} while processing the 3D volume probability maps the incorrect pixel ordering scheme was used to convert the HEALPix maps in the FITS files to the three probability density planar projection images. In this work we correct this data processing step, so that the shape of the volume projection images is consistent with the sky map and the detector network being used for each event. Furthermore, we convert the projections from probability density maps to probability maps (by multiplying by the maximum distance size scale), before normalizing. This step ensures that the projections are physically consistent and the pixels sum to unity, while not changing the information contained in the images (since they are normalized before being input to the models). We subsequently retrain the models with these updated volume images, with all other inputs remaining unchanged. We find the performance of the new updated models to be consistent with the previous models in \cite{Abbott2022}, indicating that internally the network learned the same features, or rather lack thereof, as before. This is supported by our perturbation results, which show that \gwsnm ~does not generally give preference to the volume images in making its predictions (discussed in more detail in Section \ref{subsec:learned_features}).

In this work the three one-vs-all classifiers have the same architecture and are trained on the same data set. This includes a total of 1267 glitch events identified and selected from the first two observing runs of Advanced LIGO and Advanced Virgo, as outlined in \cite{Cabero2020}. For the astrophysical CBC events a simulated set of 1000 events for each merger type of binary black hole, binary neutron star (BNS), and neutron star black hole (NSBH) is used, as outlined in \cite{Abbott2022}. This gives a total data set of 4267 events. When training each one-vs-all model, the events are labeled in a binary fashion according to whether they belong to that class or not. For example when training the BBH-vs-all model, all glitch, BNS and NSBH events are labeled as \textit{not} BBH. In this implementation of \gwsnm ~the BNS and NSBH sources are grouped into a single category of ``NS'' events that involve a neutron star in the binary merger.

\subsection{Input distributions}

\begin{figure*}
\includegraphics[width=\textwidth]{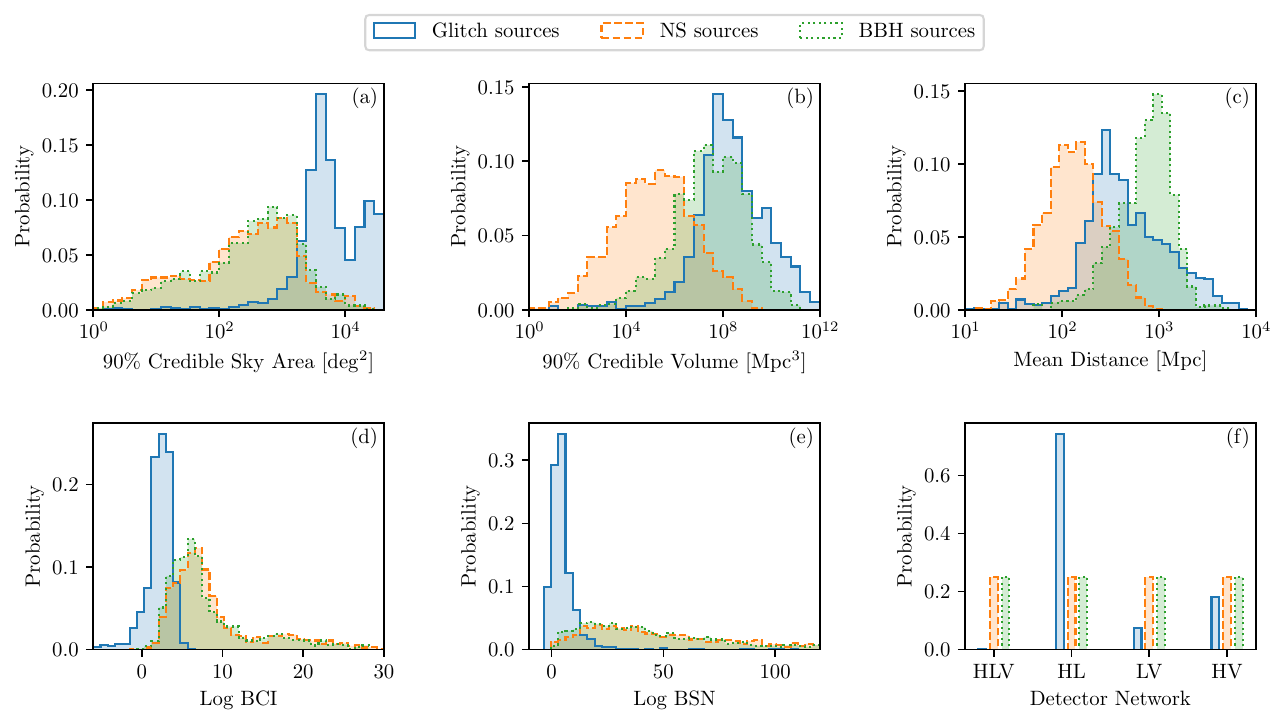}
\caption{Probability distributions of various input parameters to \gwsnm ~over the entire data set of events for the 1267 glitch (blue solid line), 2000 NS (orange dashed line) and 1000 BBH (green dotted line) sources: (a) sky map localization area, (b) 3D volume localization, (c) estimated mean distance to the source, (d) Bayes factor for coherence versus incoherence, (e) Bayes factor for signal versus noise, and (f) the network of detectors involved in the event. The distributions are distinct for certain input types and thus we expect that the models will learn to use these features to discriminate between source classes. In particular, the sky map localization area, Log BCI, and Log BSN factors can potentially distinguish between glitch events and real (NS, BBH) sources. While the mean (and maximum) distance estimate could be leveraged to classify between a merger involving a NS and a BBH merger.
\label{fig:fig1_all_inputs_original_dists}}
\end{figure*}

Before we analyze the models' predictions, we set expectations for how exactly the input features vary across the data set and for the different types of sources. This will set a baseline for determining how \gwsnm ~might be learning to distinguish the different classes. The distributions are shown in Figure~\ref{fig:fig1_all_inputs_original_dists}, aggregated by the source type for glitch, NS and BBH sources. The source distributions in our data arise from a combination of the physical limitations placed on the simulated events \citep[such as masses and spins, see][for details]{Abbott2022}, as well as the selection criteria for events, where only events with large enough signal to noise ratio (SNR $\geq 4.5$) in at least two detectors are included in the training data \citep[see][for details]{Cabero2020}.

In the sky localization area distributions in Figure~\ref{fig:fig1_all_inputs_original_dists}(a), glitch sources tend to have a wide distribution with larger values than the real NS and BBH sources, and beyond $\sim \mathrm{3000~deg^2}$ the overlap diminishes significantly. Between the NS and BBH sources the difference is not as significant, and the two populations follow very similar distributions. We thus expect that the sky map localization area will be an important feature in distinguishing between glitch and real sources in our sample.

The distribution of the 90\% credible volume in Figure~\ref{fig:fig1_all_inputs_original_dists}(b) shows how NS sources generally have a smaller volume localization than the glitch and BBH sources, reflecting the fact that they have both smaller distances (Figure~\ref{fig:fig1_all_inputs_original_dists}c) and sky localizations as compared to glitch and BBH sources. This hints that the volume projection maps might help in distinguishing sources for the NS-vs-all classifier. However, since the models only see projected images of the volume localization, the shape of the localization can still be distinct for all three sources.

For the mean distance in Figure~\ref{fig:fig1_all_inputs_original_dists}(c) it is clear that the distributions are distinct for all three sources (the distribution for maximum distance is not shown but is qualitatively the same, just scaled to larger values). Since the amplitude of the gravitational-wave signal in compact binary mergers depends on the mass of the components, in higher mass BBHs the sources can be detected to much greater distances than the lower mass mergers involving NSs. While the estimated distances for glitch sources are not physically meaningful and span a large range, they have a peaked distribution that lies between the peaks for the NS and BBH sources. We expect that the distance estimate inputs will thus be good discriminators between NS and BBH sources, but not necessarily between glitch and real sources on their own.

In Figure~\ref{fig:fig1_all_inputs_original_dists} panels (d) and (e) the computed log Bayes factors for the coherence versus incoherence hypothesis (Log BCI) and the signal versus noise hypothesis (Log BSN) show a clear distribution divide between the real sources and the glitches. The real source signals are confidently coherent across the detector network taking into account the signal time delay. For the glitch sources the values are smaller and closer to 0, but a significant fraction have Log BCIs that are positive and have some overlap with the distribution for real cases. For the Log BSN inputs, the source distributions are distinctly divided into glitch and real sources, with a significant number of glitch sources having a value greater than 0, i.e., they are distinct from background noise. Indeed this also illustrates why the positivity of Log BCI and Log BSN values on their own cannot be used to determine whether the source is a glitch or not, but must be looked at in conjunction with the other inputs. Given that the two distributions are so distinct, we expect that these Bayes factors will be leveraged heavily by the glitch-vs-all model to make accurate classifications. However, between real BBH and NS sources, these inputs may not be as discriminative.

Finally, in Figure~\ref{fig:fig1_all_inputs_original_dists}(f) the detector network for the simulated CBC sources is evenly divided in the data set between the 4 different combinations of the LIGO Hanford, LIGO Livingston, and Virgo detectors. For the glitch events, in the 2 detector configuration most glitches were detected in the Hanford-Livingston (HL) combination. This is a consequence of the fact that the two LIGO detectors were observing for most of O1 and O2, whereas the Virgo detector only joined observations for the last month of O2 \citep{Abbott2019_gwtc1}, thus leading to more accumulated glitch events in HL. In the one month period of O2 when Virgo was online, the Hanford detector was severely affected by an earthquake near the site and had a significantly lower sensitivity compared to Livingston \citep[see Figure~1 in][]{Abbott2019_gwtc1}, which lead to more glitches being identified in the HV network compared to LV during this time. There were no 3 detector HLV glitches from O2 identified in our data set. We plan to address this in future work when we re-train the models with O3 data. It is not clear how the models' classifications could be based on the detector network alone, aside from the distinction that 3 detector events will more likely be predicted as real sources instead of glitches.

\subsection{Perturbation methods}
Machine learning model explainability (and interpretability) studies can be performed using tools developed outside the domain of astronomy, for example with Gradient Weighted Class Activation Mapping \citep[Grad-CAM,][]{Selvaraju2017} for image ``attention'' mapping, Local Interpretable Model-Agnostic Explanations \citep[LIME,][]{TulioRibeiro2016} for locally approximating models, and Shapley Additive Explanations \citep[SHAP,][]{Lundberg2017} for understanding feature importance of specific events. These methods and tools can be applied to astrophysical models as appropriate to gain new insights \citep[see][for recent examples]{Wilde2022, Machado2021}. This approach was employed to perform preliminary studies of the input sky map images using Grad-CAM in \cite{Abbott2022}. In the context of the physical systems that we wish to study, explainability studies can also be performed in a simpler way by carefully examining the predictions as a function of the inputs, perturbing the inputs in systematic ways, and testing them against intuitive expectations \citep[see for example the review by][]{Ivanovs2021}.

We focus on the latter approach and quantify how the model predictions deviate as a function of the inputs and their perturbed values. Since we are interested in knowing the general features that the models are learning, this is done by training 20 different iterations of each one-vs-all model, where each iteration represents a different randomized splitting of the training and testing data set. As in \cite{Abbott2022}, this split is always performed such that 81\% of the data set is used for training, 9\% for validation, and 10\% for testing. Once the models are trained, we characterize the predictions for the events in the corresponding testing set. This is done for each input, and the results are averaged over the 20 iterations. We then systematically modify the input values in the test set one at a time and quantify how much the prediction score changes. The effects of simultaneously varying multiple input parameters that might be correlated is left to future work.

\begin{figure*}
\centering
\includegraphics[width=\textwidth]{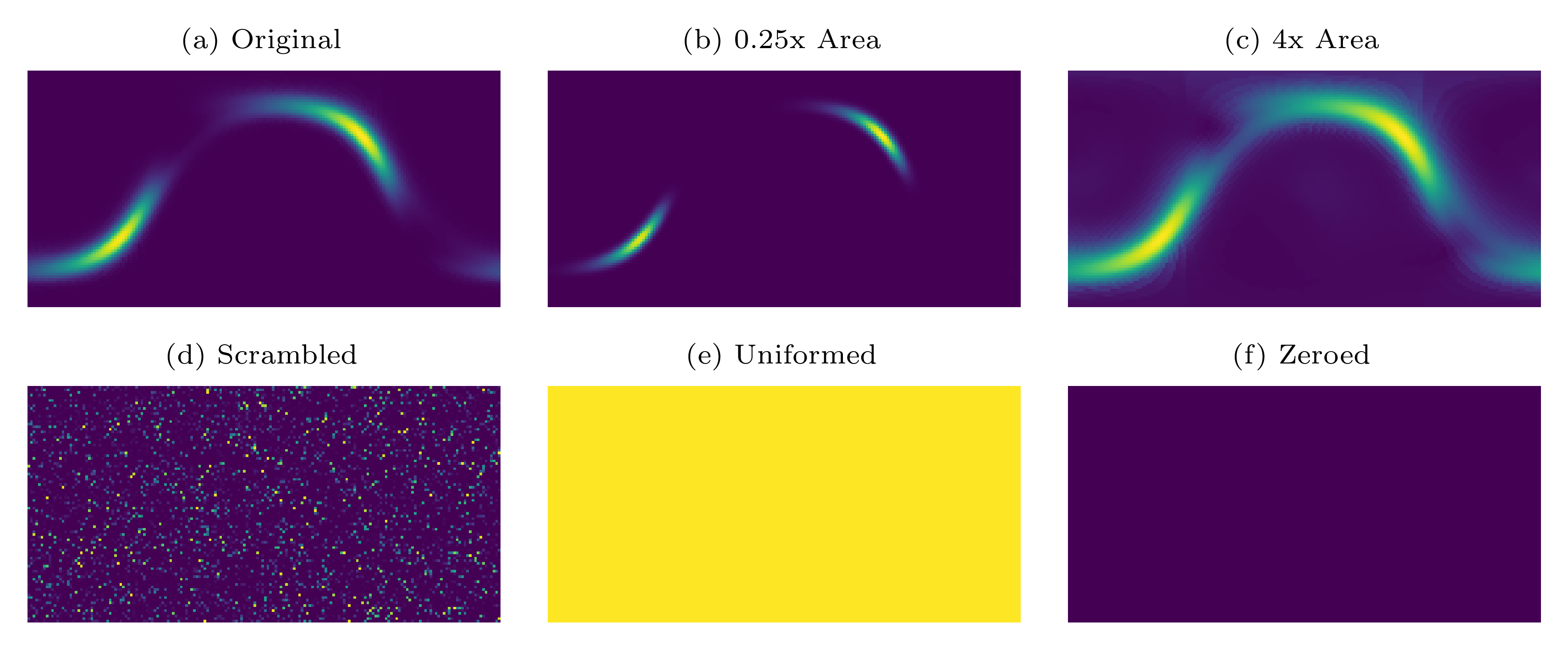}
\caption{An example of a sky map image (a) from a two-detector glitch event which has a typical 90\% sky localization area of $\mathrm{\sim 5000~deg^2}$, and the ways that it is perturbed in our studies before being input to \gwsnm: (b) down scaling the 90\% sky localization area by a factor of 4, (c) up scaling the sky localization area by a factor of 4, (d) randomly scrambling all the pixels, (e) assigning all pixels the same uniform probability, and (f) assigning all pixels a value of zero. By quantifying the effects of the perturbations on the output scores of each model in \gwsnm ~we gain insights into how it depends on these inputs and leverages the available information to perform classifications.
\label{fig:fig2_perturbed_skymaps}}
\end{figure*}

Starting with the 2D probability sky maps, we take four approaches to perturbing the data. 
\begin{enumerate}
    \item Scaling: Modify the sky map such that the 90\% credible sky localization area associated with an event is scaled by a certain factor. For example if a particular glitch event has a sky map that has a localization area of $\mathrm{5000~deg^2}$, and we scale it by a factor of 0.25, a new map is generated with the same shape and structure for the localization area, but a different size scaling such that the 90\% localization area is now $\mathrm{1250~deg^2}$, while still satisfying the normalization condition that the total sky map probability equal 1 (see Figure~\ref{fig:fig2_perturbed_skymaps}a,b). In this approach when scaling up some events with very large areas we do not allow them to exceed the $\mathrm{41253~deg^2}$ total sky area.
    \item Scrambling: Modify the sky map such that the localization area is the same, but the pixels are randomly scrambled so that there is no more identifiable shape or structure (Figure~\ref{fig:fig2_perturbed_skymaps}d).
    \item Uniforming: Make the sky maps maximally uninformative by assigning each pixel in the map the same uniform probability (Figure~\ref{fig:fig2_perturbed_skymaps}e).
    \item Zeroing: Similarly, take the extreme scenario of assigning a value of zero to each pixel, i.e., a situation where the sky map is not physically consistent with expectations and does not have any information (Figure~\ref{fig:fig2_perturbed_skymaps}f).
\end{enumerate}

The rest of the input data are perturbed in similar ways. For the 3D volume projection images, we apply the latter three approaches described above of scrambling, uniforming, and zeroing the pixels in the images. For the mean distance, maximum distance, Log BCI and Log BSN inputs we apply the first approach of scaling the original input values by certain factors, exploring a range of factors between 0.25-4. For perturbing the detector network, we cycle through the four options of having a three detector Hanford+Livingston+Virgo (HLV) detection, or a two detector HL, LV, or HV detection, i.e., we modify the detector network list wholly so that all events become one of these types. For example, if the new perturbed network is set to HLV, it means that for each event the input value for detector network is changed to [1,1,1] (regardless of what the original detector network was).

\section{Results} \label{sec:results}

\subsection{Original prediction scores}

\begin{figure*}
\centering
\includegraphics[width=\textwidth]{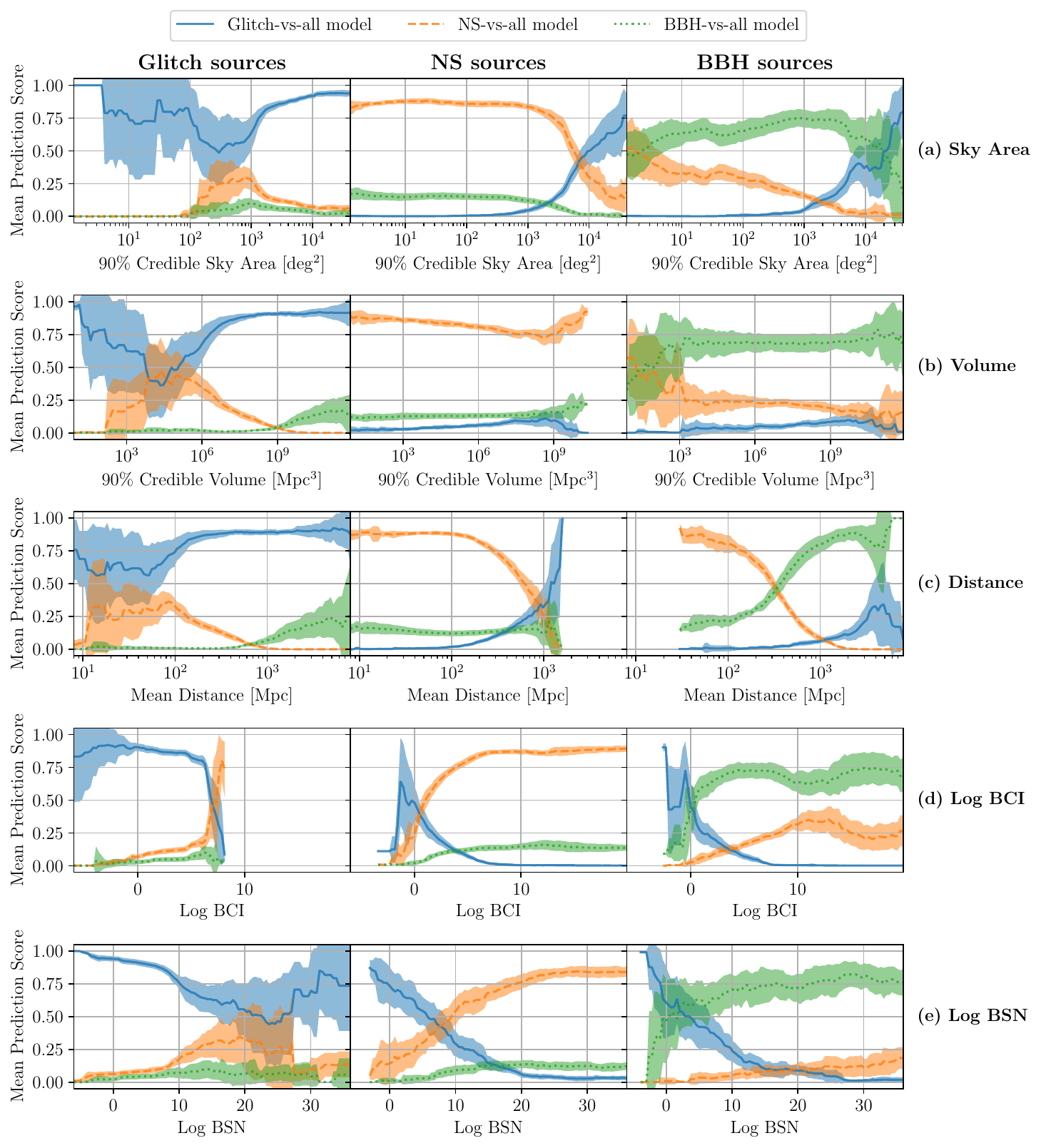}
\caption{Variation in the one-vs-all model prediction scores for the events in the testing set, as a function of select input parameters corresponding to Figure~\ref{fig:fig1_all_inputs_original_dists}. The scores are divided according to what the known source type is, where glitch sources are shown in the first column, NS sources in the second column, and BBH sources in the third column. In each panel the results for all three one-vs-all models are shown (glitch model in solid blue line, NS model in dashed orange line, BBH model in dotted green line). The prediction scores are first binned over the range of input values, and then the mean prediction score value in that bin is calculated. The shaded regions show the 1$\sigma$ standard deviation for each mean bin value over 20 trained model iterations. The results show that the models scores are primarily correlated with the sky localization area of the sky maps, the Log BCI value, and the Log BSN value in predicting whether a source is a glitch or real (NS, BBH) event. The models have also learned to associate smaller mean distances with NS events, and use this input for distinguishing between NS and BBH sources.
\label{fig:fig3_original_prediction_score_all}}
\end{figure*}

We analyze how the prediction scores in the test set change for the glitch, NS and BBH classifiers as a function of the input values, as well as whether we can see that \gwsnm ~is learning to distinguish the source classes as we would expect based on Figure~\ref{fig:fig1_all_inputs_original_dists}. These results are shown in Figure~\ref{fig:fig3_original_prediction_score_all} for the sky map localization area, the volume localization, the mean distance and the Log Bayes factors. Since we know that the prediction score will vary considerably for the different source classes, we disaggregate them according to the source type before finding the mean scores for the varying inputs.

If the source belongs to the model type (e.g., classifying glitch sources with the glitch-vs-all model), then a prediction score closer to 1 indicates that the model is correctly and confidently classifying the source. If the source does not belong to the model type (e.g., NS sources with the glitch-vs-all model) then a score closer to 0 indicates correct classification. Trends in the prediction scores, particularly in regimes where the scores for different models tend towards 0.5, signify regimes where the input parameter does not serve as a useful discriminant between the source classes and \gwsnm ~is not confident in the source classification. Where the source belongs to the model type and the score is closer to 0, the model is confidently \textit{incorrectly} classifying the source, indicating that it has learned strong dependence to associate the range of input values in that regime to one of the other source classes.

\subsubsection{Sky Localization Area}
For the sky localization area of the 2D sky maps in Figure~\ref{fig:fig3_original_prediction_score_all}(a) the general trends of the predictions are consistent with our expectations.

For glitch sources, the glitch-vs-all model confidently (with prediction scores close to 1) and accurately characterizes the event as a glitch when the sky area is large ($\gtrsim \mathrm{3000~deg^2}$), while having a larger scatter and less confidence at smaller sky areas. The NS-vs-all and BBH-vs-all models also confidently predict glitch sources to not be NS and BBH, respectively, for larger sky areas, while for smaller sky areas the score is low but not consistent with zero. There is an interesting trend at the very smallest sky areas where the NS and BBH models become more confident in their glitch classifications. This occurs because some glitch events that are loud in the detectors can produce uncharacteristically small sky maps that do not resemble astrophysical sources (e.g. the O3 candidate event S191117j\footnote{\url{https://gracedb.ligo.org/superevents/S191117j/view/}}), and thus become more distinct from real events. For example, for a two detector glitch event, if the glitch is very loud in a detector and there is some Gaussian fluctuation in the other detector that together triggers a search pipeline, then the localization error may be incredibly small --- for the event to be physically possible (astrophysical), the source has to be directly above the glitching detector to explain the large imbalance between the energy in the detectors.

For NS sources, at smaller sky areas $\lesssim \mathrm{1000~deg^2}$ all three models give the correct classifications, but at larger sky areas the predictions are less certain for the NS and glitch models and they give incorrect classifications for the very highest areas. The BBH-vs-all model becomes even more confident in correctly identifying the source as not belonging to the BBH class for the very large sky area events.

For BBH sources the prediction trends are the same as for the NS sources at the larger sky areas, but the NS-vs-all and BBH-vs-all models become less confident in their correct classifications as the sky area gets smaller, ultimately overlapping at a score of 0.5. Hence if the source is a BBH and the event sky localization is smaller than $\mathrm{\sim 5~deg^2}$, then the NS-vs-all and BBH-vs-all models are both unable to distinguish it from a NS source.

\subsubsection{Volume Localization}
Related to the sky map area trends are the volume localization trends as shown in Figure~\ref{fig:fig3_original_prediction_score_all}(b). Since the model inputs are projected maps of the volume localization which are normalized, the models do not directly access the total 90\% credible volume information. Instead, this quantity is directly related to the sky map localization area and the mean and maximum distance information.

For glitch sources with large volumes all three classifiers are able to confidently distinguish the source as a glitch (or more accurately \textit{not} NS and \textit{not} BBH for the NS and BBH models), but for lower volumes near the peak of the NS distribution of $\sim \mathrm{10^5~Mpc^3}$ (Figure~\ref{fig:fig1_all_inputs_original_dists}b), the NS-vs-all and glitch-vs-all model scores start to overlap. When the source is a NS we see that across the entire range of volumes all three models give the correct classification, with small changes in the scores across a wide range indicating that the prediction score is not strongly dependent on the volume localization value. For BBH sources the models give the correct classifications with moderate confidence for all volume localizations except for the smallest cases ($\lesssim \mathrm{10^3~Mpc^3}$), where we see the NS-vs-all and BBH-vs-all model predictions start to overlap.

\subsubsection{Distance}
The prediction trends for the mean distance are highlighted in Figure~\ref{fig:fig3_original_prediction_score_all}(c). For glitch sources the glitch-vs-all model confidently classifies the source correctly for distances $\gtrsim \mathrm{100~Mpc}$, and correctly but with less confidence for smaller distances. Since the distance estimate is not a physically meaningful quantity for glitch events, our expectation would be that the glitch-vs-all model prediction is not dependent on the distance. However, since the model has learned to associate smaller distances with NS events, it does become less confident in its prediction at large values. For the NS-vs-all and BBH-vs-all models, while the prediction score for the glitch sources is small, there is a clear trend mimicking the source distributions (as seen in Figure~\ref{fig:fig1_all_inputs_original_dists}c); for smaller distances the NS-vs-all model is not as confident that the glitch is not a NS, and for larger distances the BBH-vs-all model is not as confident that the glitch is not a BBH.

For NS sources the NS-vs-all model score is highly dependent on the distance. The model is confident and accurate at smaller distances, but beyond $\sim 500$ Mpc the classification changes and the NS-vs-all model does not identify the NS sources. The BBH-vs-all model score stays consistent at a smaller distance in this regime, but the glitch-vs-all score actually increases with distance, indicating a learned inclination to classify NS sources at larger distances as glitches instead. For the BBH sources the NS-vs-all model score is actually higher than the BBH-vs-all model score for distances below $\sim 500$ Mpc. Both the NS-vs-all and BBH-vs-all models associate smaller distances with NS sources and larger distances with BBH sources.

\subsubsection{Bayes factors}
For the dependence on the coherence versus incoherence Bayes factor, the results in Figure~\ref{fig:fig3_original_prediction_score_all}(d) are consistent across all three source types for each model. They show that all three models have learned that Log BCI values less than 0 are associated with glitch sources, even if the source may actually be real. However, there is a sharp transition near the 0 point for real sources beyond which they are correctly classified by all three models. For glitch sources, this transition actually occurs closer to a value of Log BCI $\sim 7$, indicating that even if the glitch event has marginal support for the coherent hypothesis over incoherent across the two detectors, it will still tend to be correctly classified as a glitch unless the log Bayes factor reaches an overwhelming threshold value of 7 or more, strongly favoring coherence.

The trends for the signal versus noise Bayes factors in Figure~\ref{fig:fig3_original_prediction_score_all}(e) are qualitatively similar to that of Log BCI, except they do not show as sharp a transition between prediction scores for glitch sources. In fact for glitch sources the glitch-vs-all model score varies considerably for larger Log BSN values, with mean prediction values around $\sim 0.6$. For the real NS and BBH sources, the model prediction scores confidently classify the sources incorrectly as glitches for smaller Log BSN values, but correctly for larger values. This suggests that all models have learned to associate smaller Log BSN values with glitches (or a correlated quantity, as becomes apparent in the perturbation results).

\subsection{Perturbation effects}

\begin{figure*}
\includegraphics[width=\textwidth]{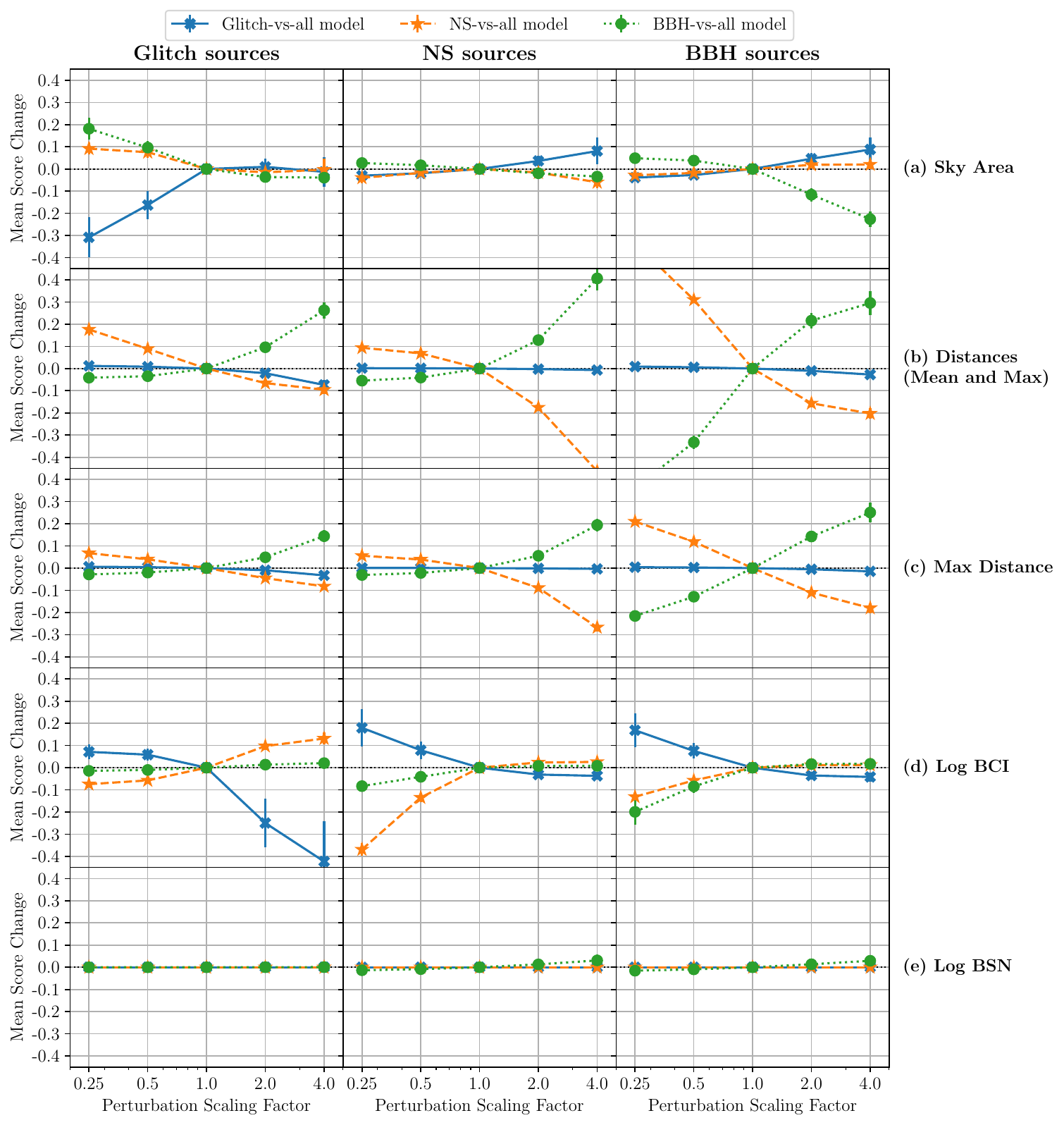}
\caption{The effects on the model prediction scores of perturbing (scaling) select input data: (a) 90\% sky localization area, (b) mean and maximum distance estimates simultaneously, (c) only maximum distance estimate, keeping mean distance the same, (d) coherence versus incoherence Bayes factor, and (e) signal versus noise Bayes factor. The columns show the results divided according to the known source type, and each panel includes the individual results for all three models (glitch model in blue crosses and solid line, NS model in orange stars and dashed line, BBH model in green circles and dotted line). Each row corresponds to perturbing a single input type. The original values are varied by a factor of 0.25 to 4, and the change in the prediction score is calculated using the corresponding prediction score for the unperturbed inputs. The mean value of the score change is computed for each scaling factor, and the 1$\sigma$ standard deviation is calculated over the results of the 20 trained model iterations. The trends show that increasing the distances produces a large change in the prediction score in favor of classifying the source as a BBH as opposed to an NS, while it has a minimal effect on the output of the glitch-vs-all model. Increasing the sky localization area and decreasing the Log BCI factor are associated with an increased propensity to classify a real source as a glitch. The prediction score is not sensitive to perturbations of the Log BSN value for any of the three models, indicating that \gwsnm ~actually does not consider this input important.
\label{fig:fig4_prediction_score_change_scalar_all}}
\end{figure*}

\begin{figure*}
     \centering
     \includegraphics[width=\textwidth]{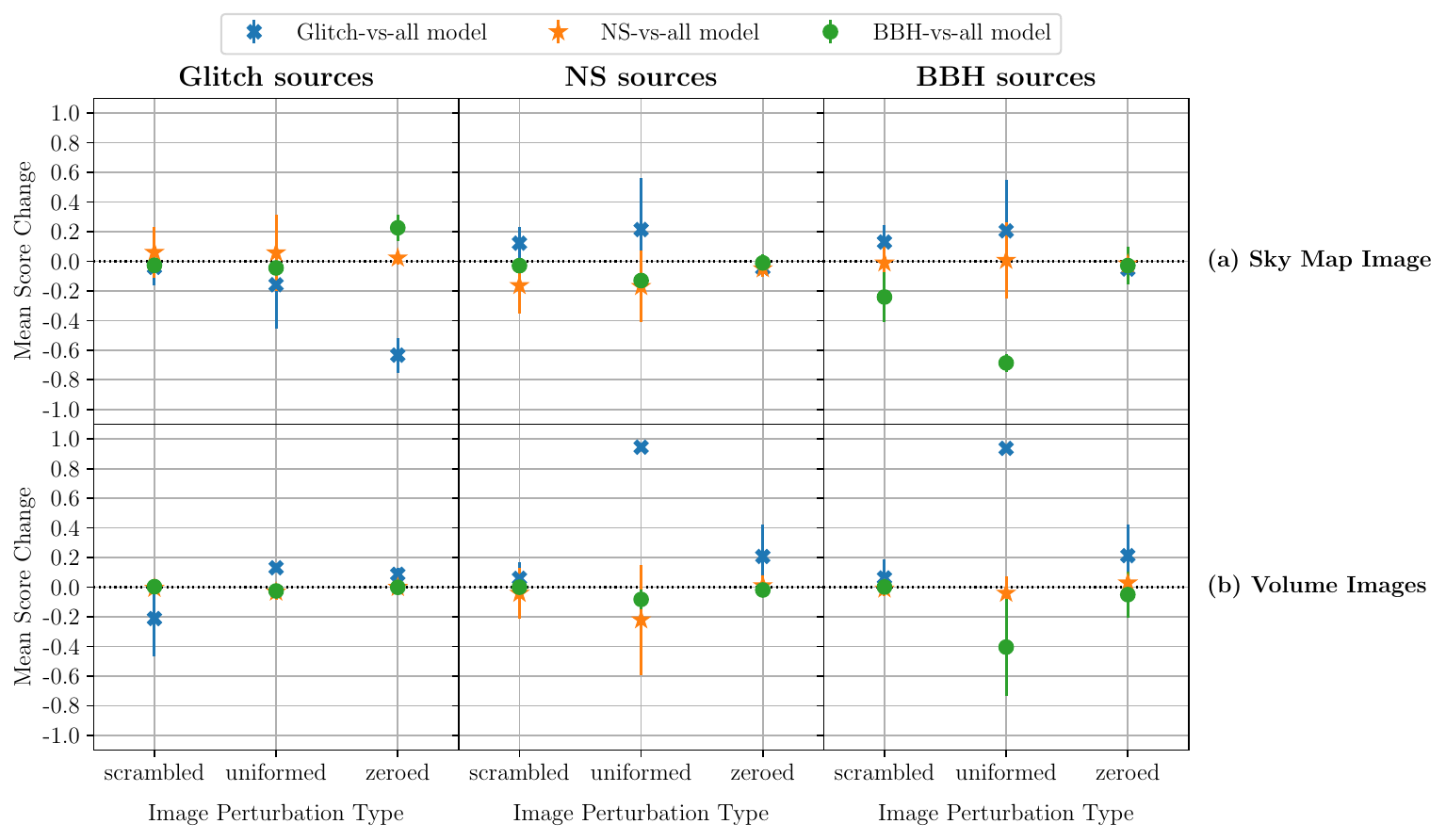}

     \bigskip
     \includegraphics[width=\textwidth]{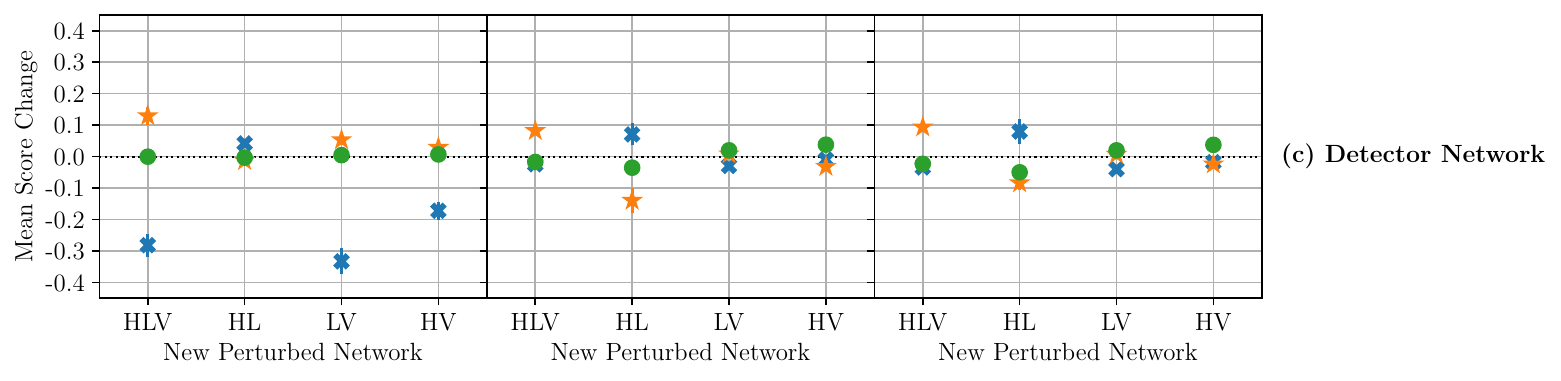}
    \caption{The effects on the model prediction scores of perturbing the input data, following the same layout as in Figure~\ref{fig:fig4_prediction_score_change_scalar_all}, but for non-scaling perturbations: (a) manipulating the pixels in the sky map images, (b) manipulating the pixels in the volume projection images, and (c) changing the detector network to an entirely new value. The changes in the scores for both the sky map images and volume images indicate that the models look for an overall shape and structure that is consistent with the other inputs, but are largely driven by the size of the localization; the predictions favor glitch classifications when there is no identifiable shape or the images are uniform (large areas/volumes). The BBH-vs-all model relies more on the images to make the correct classification than the NS-vs-all classifier. The score changes in the detector network show that the BBH-vs-all model is insensitive to this information, while the NS-vs-all model uses three detector events as support for the source being a NS merger. The glitch-vs-all model changes follow the same pattern as seen in the original input distribution in Figure~\ref{fig:fig1_all_inputs_original_dists}(f), with HL events being associated more with glitches than events involving the Virgo detector.}
    \label{fig:fig5_prediction_score_change_nonscalar_all}
\end{figure*}

We show the systematic effects of perturbing our inputs on the output prediction scores in Figure~\ref{fig:fig4_prediction_score_change_scalar_all} for the scaled values, and in Figure~\ref{fig:fig5_prediction_score_change_nonscalar_all} for the modifications to the image-based inputs as well as the detector network changes. A complementary approach to quantifying these perturbation effects using the overall model accuracy score instead of the mean prediction score is included in Appendix \ref{sec:appA}.

\subsubsection{Sky Localization Area}
We first quantify the effects of varying the sky map localization areas for all classifiers. We see in Figure~\ref{fig:fig4_prediction_score_change_scalar_all}(a) that overall the models have learned to associate larger sky areas with glitches, and smaller sky areas with real events, but the effect is not as large as it is for distances (Figure~\ref{fig:fig4_prediction_score_change_scalar_all}b). 

If the source is a glitch, then increasing the sky area results in a small decrease in the BBH-vs-all score, while the glitch and NS model results stay the same. However, when decreasing the sky area the effect is larger, with all three models having a higher likelihood of misclassifying the source: the glitch-vs-all model score decreases while the NS and BBH model scores increase. These results make sense in the context of the sky area distributions shown in Figure~\ref{fig:fig1_all_inputs_original_dists}(a). Most NS and BBH have localization areas below $\sim \mathrm{3000~deg^2}$, while most glitches have areas above this value, and the distributions are distinct for real vs glitch events. When the glitch sky areas are decreased they have more overlap with the NS and BBH regime, whereas when they are increased there is minimal effect since the glitch sky areas are already much larger.

Increasing the sky area has the same effect of increasing the glitch-vs-all score for the NS and BBH sources by $\sim 0.1$. However, the change in the BBH-vs-all score for BBH sources is much larger than the change in NS-vs-all score for NS sources, especially when increasing the sky area, indicating that the BBH-vs-all model has learned to give more weight to sky area than the NS-vs-all model.

\subsubsection{Distance}
Similarly, we consider the effects of varying the source distance for all classifiers. Since the models take both the mean and maximum distance estimates as inputs, we show the effects of scaling both values at the same time by the same factor (Figure~\ref{fig:fig4_prediction_score_change_scalar_all}b), and of only varying the maximum distance while keeping the mean distance constant (Figure~\ref{fig:fig4_prediction_score_change_scalar_all}c).

We observe that the glitch-vs-all model is relatively unaffected by changes in distance, while NS and BBH predictions vary monotonically. There is an exception when the distance is increased by 4 times for glitch sources; here the model score decreases as it associates these much larger distances with BBH sources. The glitch classifier has learned that the distance estimate cannot be used to distinguish between real events and glitches, as the glitch distance distribution in Figure~\ref{fig:fig1_all_inputs_original_dists}(c) is very similar to the combined real event distribution (NS+BBH). This suggests that the higher score for the glitch-vs-all model seen with larger distances in Figure~\ref{fig:fig4_prediction_score_change_scalar_all}(c) is likely because the higher distance glitch events are correlated with larger sky areas and/or smaller Log BCI values.

The NS and BBH models, however, learn that the distance is proportional to source mass for real events, as higher mass BBH events are identified by the BBH model with larger distances, and lower mass NS merger events are identified by the NS model with smaller distances. That is, if the source is a NS or BBH, all three classifiers are learning the correct distance dependence (and independence) and accounting for it in the predictions. If the source is a glitch, then i) the glitch classifier is (almost) independent of distance, and ii) NS and BBH classifier predictions are affected by distance.

A subtle point is that when the model prediction score decreases, it only signifies that the model is less likely to classify the source as \textit{not} belonging to that class, but on its own does not say anything about which of the other two classes it is more likely to be. For example for the NS-vs-all model, increasing the distance decreases the prediction score, which means the model is less likely to classify the source as a NS, but does not indicate whether it is more likely to be a BBH or a glitch. However, when we consider this along with the observations that the BBH model score increases while the glitch model score stays the same, we can break this degeneracy and find that sources with larger distance estimates are more likely to be classified as BBH instead of NS.

\subsubsection{Bayes Factors}
The results of perturbing the Log BCI input are shown in Figure~\ref{fig:fig4_prediction_score_change_scalar_all}(d). While most Log BCI values for events are positive, in principle because this input spans both positive and negative values for glitch sources, perturbing the value with a scaling factor $> 1$ means that it is further away from zero and more confidently in support of event coherence (or incoherence) across detectors, while perturbing the value with a scaling factor $< 1$ means that it is closer to zero and there is less support than before for preferring one of the coherence or incoherence models over another.

Figure~\ref{fig:fig4_prediction_score_change_scalar_all}(d) shows that the changes in scores are particularly significant when the Log BCI is scaled up for glitch sources and when it is scaled down for real sources. For glitch sources the glitch-vs-all classifier score decreases by $\sim 0.4$ when it is scaled up by a factor of 4, resulting in many of these glitches being misclassified as real events. In contrast, the score increases slightly when scaling the Log BCI values down. This indicates that for Log BCI values further away from zero the glitch-vs-all model becomes less confident that the event is a glitch. While the BBH-vs-all model score remains virtually unaffected for the glitch sources, the NS-vs-all model score changes by up to $\sim \pm 0.1$, indicating a larger dependence on the Log BCI factor for the NS-vs-all model as compared to the BBH-vs-all model. When the Log BCI value is increased for the NS and BBH sources there is an insignificant change in the NS-vs-all and BBH-vs-all model scores, and a slight decrease in the glitch-vs-all model score. The NS and BBH sources mostly have large Log BCI values, as shown in the distribution in Figure~\ref{fig:fig1_all_inputs_original_dists}(d), so this result is in line with our expectation that increasing the values further will have minimal effect. When decreasing the Log BCI value the NS-vs-all and BBH-vs-all prediction score should decrease, which is precisely what we observe in the perturbation effects. When comparing the size of the prediction score decrease for the NS-vs-all model for NS sources ($\sim -0.4$ for a factor of 0.25) and the BBH-vs-all model for BBH sources ($\sim -0.2$ for a factor of 0.25), we again see that the effect of perturbing the Log BCI value on the NS-vs-all model is larger than on the BBH-vs-all model.

For perturbing the Log BSN factor Figure~\ref{fig:fig4_prediction_score_change_scalar_all}(e) shows that besides a very small increasing trend for the BBH-vs-all model score for the NS and BBH sources, there is no change in the scores for any of the models for any of the sources. This means that the Log BSN factor is given virtually no weight in any of the models and is not being used by \gwsnm ~to perform any classification decisions. To see if any change in score would arise for more extreme perturbations, we show in Figure~\ref{fig:fig10_prediction_score_change_logBSN_extended} in Appendix~\ref{sec:appB} the results for perturbing the Log BSN values by factors of up to 16 times larger or smaller. Even then the results are the same, with the slight dependence of the BBH-vs-all classifier for the real sources visible, but no changes otherwise, confirming that the \gwsnm ~predictions are not sensitive to the Log BSN input value.

\subsubsection{Sky Map}
For the non-scaling perturbations, Figure~\ref{fig:fig5_prediction_score_change_nonscalar_all}(a) shows how the prediction scores change when we manipulate the pixels in the sky map images (scrambling, uniforming, and zeroing).

When the pixels are scrambled, for glitch sources the score does not change significantly for any of the models, indicating that for these events the models do not focus on the specific distribution of the pixels in the sky maps. For the real NS and BBH sources, there is also a small increase in the glitch-vs-all model score of $\sim 0.1$, which means that when we remove structure from the real event sky maps the glitch-vs-all model has a slightly higher tendency to classify those events as glitches instead. Furthermore for NS sources the NS-vs-all model score decreases while the BBH-vs-all model score stays the same (and vice-versa for BBH sources). The BBH-vs-all model has thus learned to ignore the shape or structure of the localization sky map for sources that are not BBH (and the NS-vs-all model for sources that are not NS).

For the uniformed sky maps there does seem to be an effect on the models, but there are also large uncertainties overlapping with the zero point indicating that these effects are not consistent. The exception is that the BBH-vs-all model score decreases significantly by $\sim 0.7$ for BBH sources, which means that the model heavily weighs the size and shape of the sky map for BBH sources, misclassifying the events as not BBH if the sky map is not consistent with the other inputs.

When the sky map image pixels are zeroed there is no change in any of the model scores if the source is a NS or BBH, but the score significantly decreases by $\sim 0.6$ for the glitch-vs-all model and increases by $\sim 0.2$ for the BBH-vs-all model when the source is a glitch. The NS-vs-all model has thus learned to ignore the sky map if it has no information, but the BBH-vs-all and glitch-vs-all models have a tendency to correlate these empty sky maps with real sources.

\subsubsection{Volume Images}
The effects of perturbing the three volume projection images can be seen in Figure~\ref{fig:fig5_prediction_score_change_nonscalar_all}(b).

Scrambling the pixels in the images does not have any effect on the output of the NS-vs-all and BBH-vs-all models, and only a marginal effect on the glitch-vs-all model for glitch sources only, where the score slightly decreases. The models are thus not utilizing the projected shape of the localization to inform prediction scores.

When the volume images are uniformed, for glitch sources the NS-vs-all and BBH-vs-all model scores are unaffected, while the glitch-vs-all model score increases by $\sim 0.1$ so that it is slightly more confident in making the correct classification. For NS and BBH sources, we see that the change in results for the glitch-vs-all model are extreme and the model associates uniform volume maps exclusively with glitches. There is a decrease in the NS-vs-all score for NS sources and in the BBH-vs-all score for BBH sources, indicating again that these are being misclassified as glitches for uniform volume images, but the associated uncertainties are also very large.

Similarly, zeroing the pixels in the volume images does not affect the NS-vs-all and BBH-vs-all model predictions, but does marginally increase the glitch-vs-all model score for real sources, indicating that the glitch-vs-all models have learned a marginal bias towards classifying real sources as glitches if the volume maps are empty.

\subsubsection{Detector Network}
Figure~\ref{fig:fig5_prediction_score_change_nonscalar_all}(c) shows how the model prediction scores change when the models are given new perturbed inputs for the detector network.

A three-detector Hanford-Livingston-Virgo (HLV) network increases the NS-vs-all score by $\sim 0.1$ for all sources, which means it has learned to associate HLV events with a higher probability that the source is a NS. The same is not true for the BBH-vs-all model, which remains unaffected when any of the two detector events are perturbed to be HLV events. Thus it is not that the models predict any real event to be more likely if it is a three detector event, but rather the NS-vs-all model favors classifying three detector events as being NS mergers. For the HLV network the glitch-vs-all model score is unaffected if the source is real, but decreases significantly if it is a glitch, reflecting the fact that there are no three detector glitch events in our data set (see Figure~\ref{fig:fig1_all_inputs_original_dists}f).

If the detector network is perturbed from its original value to be a two detector event (HL, LV, or HV), then the effects on the models are varied and largely follow their respective original source distributions as shown in Figure~\ref{fig:fig1_all_inputs_original_dists}(f). The BBH-vs-all model scores are unchanged for glitch sources, and insignificantly affected for NS and BBH sources. The NS-vs-all model scores only decrease by $\sim 0.1$ for NS and BBH sources when the new perturbed network is HL, but not otherwise, whereas the glitch-vs-all model scores increase by $\sim 0.1$ for these same cases. HL events are thus associated with an increased probability of being glitches, mirroring the large number of glitch events seen in this two detector configuration in Figure~\ref{fig:fig1_all_inputs_original_dists}(f). For glitch sources we similarly see the glitch-vs-all model scores decrease if the input network is changed to LV or HV, i.e. non-HL events are associated with a lower likelihood of being a glitch.

\section{Discussion} \label{sec:discussion}

\subsection{Learned input features and trends} \label{subsec:learned_features}
Our results show how \gwsnm ~relies on its input parameters to accurately classify the source of LIGO-Virgo gravitational-wave events. The effects of perturbations to the inputs (Figures \ref{fig:fig4_prediction_score_change_scalar_all} and \ref{fig:fig5_prediction_score_change_nonscalar_all}), when put in the context of the original predicted trends (Figure~\ref{fig:fig3_original_prediction_score_all}) and the input distributions (Figure~\ref{fig:fig1_all_inputs_original_dists}), paint a coherent picture of how each of the three models in \gwsnm ~use the inputs to distinguish glitch, NS and BBH sources.

For the sky map images including the localization area, \gwsnm ~looks at how large the localization area is and predicts that sources with larger areas are glitches instead of real events. Intuitively this is consistent with our expectations; glitch events are not physically correlated across different detectors, but occur due to chance alignment of transient noise, and so are harder to coherently model by \texttt{BAYESTAR}, resulting in wider posterior distributions of the sky location parameters and thus larger sky localization areas. This sky area dependence is seen for all three models. However, the change in the BBH-vs-all scores for perturbations are larger than the change in NS-vs-all scores, indicating that the BBH-vs-all model has learned to give more weight to sky area than the NS-vs-all model. The glitch-vs-all model has learned thresholded sky area dependence: below a certain sky area ($\lesssim \mathrm{3000~deg^2}$) it associates events more with NS and BBH sources. Perturbing the sky map images confirms this and shows that the models look for the presence of an overall shape and structure that is consistent with the other inputs, but is largely driven by the size of the localization. The predictions favor glitch classifications when there is no identifiable shape or the images are uniform (large area). Here again the BBH-vs-all model relies more on the images to make the correct classification than the NS-vs-all model.

The impact of the three projected volume localization images is more pronounced on the glitch-vs-all model as compared to the NS-vs-all and BBH-vs-all models. For the latter two cases the models are not utilizing the specific projected shape of the localization to inform prediction scores and only show a very marginal dependence on the projected localization sizes. When the source is a NS all three models give the correct classification, and for BBH sources the models give the correct classifications with moderate confidence for all volume localizations except for the smallest case. When these are perturbed, the glitch-vs-all model is the only one that is impacted as it correlates the large projected volumes in the uniformed images with glitch sources. Thus the volume images are only leveraged by the glitch-vs-all model in a meaningful way.

Studying the mean and maximum distance estimates shows that they are a very important input to the \gwsnm ~NS-vs-all and BBH-vs-all models in distinguishing between the real NS and BBH sources, but the predictions for the glitch-vs-all model are less impacted by this input. Since the distance estimate is not a physically meaningful quantity for glitch events, this is in line with our expectations that the glitch-vs-all model prediction is largely independent of distance. For the real sources, the perturbation effects show that increasing the distances produces a large change in the prediction score in favor of classifying the source as a BBH as opposed to a NS, while smaller distances favor classifying the source as a NS. This directly follows from the distribution of source distances in our data set (Figure~\ref{fig:fig1_all_inputs_original_dists}(c)).

While having a class balanced approach in the data set facilitates machine learning model training, in reality the number of NS events detected by the LVK is much smaller than the number of BBH events: 3 NS compared to 40 BBH in O3 \citep{Abbott2021_gwtc3}. This is expected since the amplitude of the GW signal is proportional to the source mass, and so BBH can be detected to much larger distances. Thus \gwsnm ~learning to associate larger distance events with BBH is the physically correct learned behavior. But the converse is not strictly true: in reality an event detected at a smaller distance could be a BBH or NS merger. Considering the O3 real events from GWTC-3, for example, 3/40 BBH and 3/3 NS events were detected at relatively smaller distances $\mathrm{\lesssim 500 ~Mpc}$ \citep{Abbott2021_gwtc3}. Thus the models predicting smaller distance events to more likely be NS instead of BBH shows a bias of \gwsnm ~based on the training set that is not physically true. This is a trade-off for having a class balanced number of events in the data set to facilitate accurate model training, even though in reality the detected event sources are heavily imbalanced.

We also show that \gwsnm ~has learned to use the Log BCI factor to distinguish between real sources and glitch sources. Even if a glitch event seemingly has support for being coherent across the two detectors with Log BCI $> 0$ it will still tend to be correctly classified as a glitch unless the Bayes factor starts to reach large values of beyond $\sim 7$. Decreasing the Log BCI factor is associated with an increased likelihood to classify a real source as a glitch. There is a larger dependence on the Log BCI factor for the NS-vs-all model as compared to the BBH-vs-all model: \gwsnm ~has learned that the degree of coherence across detectors is a more important attribute to classify NS sources than BBH sources.

The NS-vs-all model thus gives larger weight to the Log BCI factor as compared to the BBH-vs-all model. We can contrast this to the BBH-vs-all model having a higher dependence on the sky map images. The BBH-vs-all model thus focuses more on coherence information that is encoded in the shape of the sky map image, as opposed to the value of the Log BCI factor directly.

The effects of perturbing the Log BSN value show a rather unexpected result: there is virtually no change in the prediction scores. This means that the Log BSN factor is given no weight in any of the models and is not being used by \gwsnm ~to perform any classification decisions. The result is contrary to our expectations since the distributions in Figure~\ref{fig:fig1_all_inputs_original_dists}(e) suggest the models would learn to use this value to distinguish between glitch and real events, and the trends in Figure~\ref{fig:fig3_original_prediction_score_all}(e) indicate that all models are indeed learning to associate smaller Log BSN values with glitches, and larger values with NS and BBH sources. Our perturbation results indicate that the trends we see in Figure~\ref{fig:fig3_original_prediction_score_all}(e) are due to correlation and not causation --- larger Log BSN values correlate with larger Log BCI values and smaller sky areas (see Figure~\ref{fig:fig11_BSN_BCI_correlation_scatter} in Appendix~\ref{sec:appB}), and it is the latter two inputs that are important in distinguishing glitch and real sources. It is not clear why the Log BSN factor is not being used. One possible explanation is that \gwsnm ~is using the sky area and coherence information to independently infer signal versus noise information in a form that is more useful in distinguishing real versus glitch events, and so it ignores the external input Log BSN value entirely. We leave a more thorough investigation of this to future work.

Finally, perturbing the detector network information shows that \gwsnm ~has learned to associate two detector HL events with an increased probability of being glitches, consistent with the large number of glitch sources seen in this two detector configuration in Figure~\ref{fig:fig1_all_inputs_original_dists}(f). We also see that the NS-vs-all and BBH-vs-all models have learned a different dependence on three detector events versus two detector events. The NS-vs-all model associates three detector HLV events with NS sources more strongly, whereas the BBH-vs-all model remains largely unchanged. This again points to our earlier remark that the BBH-vs-all model is finding and utilizing coherence information more from the shape of the sky map rather than from the Log BCI value or the detector network.

\subsection{Understanding O3 event misclassifications}\label{subsec:O3_misclass}
Our perturbation results and insights into what the models have learned can also be used to understand the events that are misclassified by \gwsnm ~in the LVK third observing run (O3).

For each CBC candidate event in O3 for which an OPA was issued, the predicted class from \gwsnm ~is compared to the final LVK classification of the event in GWTC-3 \citep{Abbott2021_gwtc3}. In our analysis only events with $p_{\mathrm{astro}} > 0.5$ in GWTC-3 are considered as real astrophysical events, with the rest classified as glitches (the marginal NSBH events S190426c with $p_{\mathrm{astro}} = 0.14$ and S200105ae with $p_{\mathrm{astro}} = 0.36$ are thus classified as glitches in our analysis). The type of binary merger for the real events (BNS, NSBH or BBH) is then determined from the given best-fit component masses, assuming a maximum NS mass of $\mathrm{3~M_{\odot}}$. For the \gwsnm ~predictions the three glitch-vs-all, NS-vs-all, and BBH-vs-all models are used in a hierarchical scheme to produce a single classification for each event \citep[as in][]{Abbott2022}. The prediction scores and classifications are updated in this work by using the average results of the 20 trained model iterations, as compared to the result of 1 selected model iteration in \cite{Abbott2022}. This helps to reduce variance in the predictions and produce a classification that is more robust to model outliers. The averaging results in 5 events (out of 77) in O3 having a different final classification.

\begin{figure}
\includegraphics[width=\columnwidth]{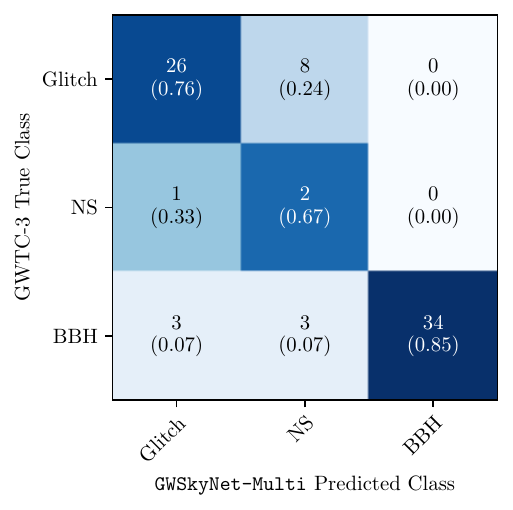}
\caption{Predicted versus true classification confusion matrix for the 77 CBC candidate events in O3 for which an OPA was issued. The predicted classifications are the \gwsnm ~hierarchical ones. The true classifications are taken to be the ones from the final LVK O3 event catalog GWTC-3 \citep{Abbott2021_gwtc3}. \gwsnm ~misclassifies 15 (20\%) of these events, listed in Table ~\ref{table:O3_misclass} in Appendix~\ref{sec:appC}. 
\label{fig:fig6_O3_classifications_matrix}}
\end{figure}

Figure~\ref{fig:fig6_O3_classifications_matrix} shows the confusion matrix for the predicted classifications of the 77 O3 candidate events. \gwsnm ~misclassifies 15 (20\%) of these events (8/40 in O3a and 7/37 in O3b), listed in Table~\ref{table:O3_misclass} in Appendix~\ref{sec:appC}. All glitch events that are misclassified by \gwsnm ~are identified as NS merger events, while no glitch or NS event is misclassified as a BBH (i.e., there are 0 BBH false positives).

\begin{figure*}
\includegraphics[width=\textwidth]{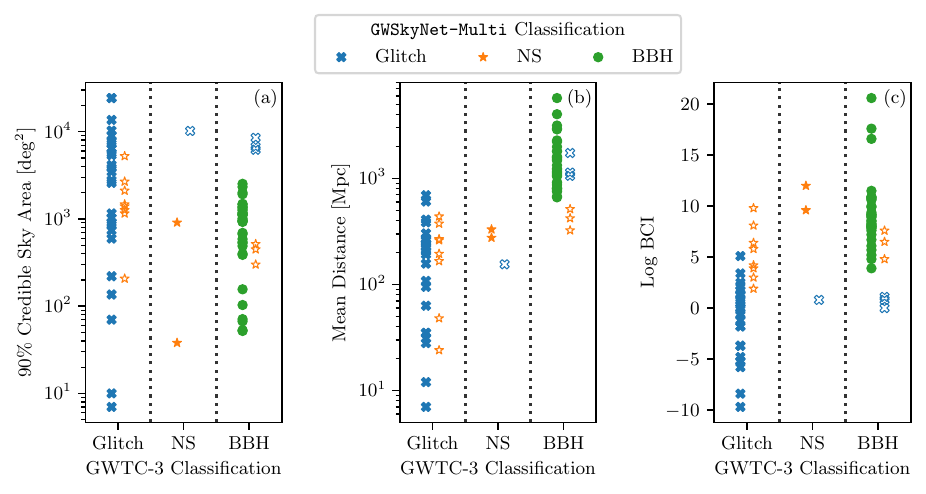}
\caption{Classifications of the 77 CBC candidate events in O3 for which an OPA was issued, distributed vertically according to (a) the sky map localization area, (b) estimated mean distance, and (c) Bayes factor for coherence versus incoherence. In each panel the events are horizontally divided according to their true classification from GWTC-3: 34 glitch events in the left column, 3 NS merger events in the center column, and 40 BBH merger events in the right column. The marker for each event is labeled according to its predicted classification from \gwsnm: glitches in blue crosses, NS in orange stars, and BBH in green circles. Events that are correctly classified by \gwsnm ~have filled markers (e.g., the filled blue crosses for glitches in the glitch column), while misclassified events have open markers that are slightly offset to the right (e.g., the unfilled orange stars for NS in the glitch column). The misclassified events can be compared to the correct classifications and understood in the context of our perturbation study results. We find that the 4 real events misclassified as glitches have larger sky areas and smaller Log BCI values than the rest of the real event populations. Complementarily the 8 glitch events misclassified as NS have smaller sky areas and larger Log BCI values. The 3 BBH misclassified as NS have smaller distances.
\label{fig:fig7_O3_classifications_vs_input_all}}
\end{figure*}

The misclassified events can be understood in the context of our perturbation study results when we look at the sky localization area, distance, Log BCI and detector network that are input to \gwsnm ~from the associated \texttt{BAYESTAR} FITS files released in the OPAs. Figure~\ref{fig:fig7_O3_classifications_vs_input_all} shows some of these input values for the 77 events, which are divided according to their true source classification, with events misclassified by \gwsnm ~highlighted.

In Figure~\ref{fig:fig7_O3_classifications_vs_input_all}(a) the 3 BBH and 1 NS merger events misclassified as glitches all have very large sky areas $\mathrm{\gtrsim 5000 ~deg^2}$, which is distinct from the correctly identified NS and BBH events all of which have smaller sky areas $\mathrm{\lesssim 3000 ~deg^2}$.  Conversely, aside from S191220af, all glitches classified as NS have sky areas $\mathrm{\lesssim 3000 ~deg^2}$. These misclassifications align with our result that \gwsnm ~focuses on the sky localization area in the sky maps to distinguish between glitch and real sources, and tends to predict sources with large areas to be glitches instead of real events.

In Figure~\ref{fig:fig7_O3_classifications_vs_input_all}(b) the 3 BBH events misclassified as NS all have the smallest estimated mean distances within the BBH population, with values $\mathrm{\lesssim 500 ~Mpc}$. This is consistent with our expectations, as \gwsnm ~has been shown to use the distance estimate as a strong discriminator between BBH and NS events, favoring classifying the source as a NS merger when the distance is small.

Figure~\ref{fig:fig7_O3_classifications_vs_input_all}(c) highlights the classifications against the input Log BCI value, wherein the 3 BBH and 1 NS merger events misclassified as glitches all have small Log BCI values $\sim 0-1$ (virtually no support for the coherence hypothesis versus incoherence), while the rest of the NS and BBH events have larger Log BCI values. Conversely, all 8 glitch events misclassified as NS have Log BCI $\gtrsim 2$, with 4/8 having Log BCI $\gtrsim 6$, significantly larger than the rest of the glitch events. These trends are consistent with our perturbation results, which show that \gwsnm ~considers the Log BCI value an important parameter in deciding between real and glitch events, decreasing the glitch score for larger Log BCI values and classifying the events as real. Furthermore, as the NS-vs-all model gives larger weight to the Log BCI value as compared to the BBH-vs-all model, these glitch events with large Log BCI tend to be misclassified as NS events as opposed to BBH.

A look at the detector network for the O3 event misclassifications reinforces the need to update the training set for \gwsnm ~with more glitches involving the Virgo instrument. There are a total of 32 2-detector events in O3 --- 19 HL, 11 LV, and 2 HV. Of these only 2/19 (11\%) HL events are misclassified by \gwsnm, but when the Virgo detector is observing the misclassification rate goes up significantly, with 6/11 LV events and 2/2 HV events (total 8/13 or 62\%) misclassified. In sum, 8/10 2-detector misclassifications involve Virgo. This likely stems from the fact that the Virgo instrument is less sensitive than Hanford and Livingston: while it may be observing during an event, the SNR in Virgo is likely much lower than in HL, effectively making it a single detector event, which \gwsnm ~is not trained to predict on. For example the event S200302c is listed as a HV event on GraceDB\footnote{\url{https://gracedb.ligo.org/superevents/S200302c/view/}}, but the individual detector SNRs from GWTC-3 are 10.4 in Hanford and 1.9 in Virgo \citep{Abbott2021_gwtc3}, which means the event would be annotated as a single detector Hanford event according to our training set criteria (require SNR $\geq 4.5$ in at least two detectors). In future work we will update the training set with O3 glitches involving Virgo, and explore adjusting the SNR threshold for selecting events to reduce this apparent discrepancy between the detectors in the network that are online during an event and the detectors that actually observe an associated signal with high enough SNR.

\section{Conclusion} \label{sec:conclusion}
Deep learning models have been applied successfully in the physical sciences and continue to increase in popularity. The trade-off for increase in performance in this domain is the associated increase in complexity, which makes it challenging to understand what the model has learned and what its limitations are.

In this paper we present our work to understand one such model for classifying LIGO-Virgo gravitational-wave events: \gwsnm. To increase our trust in the model and verify its robustness, we study in detail how the three underlying glitch-vs-all, NS-vs-all and BBH-vs-all models behave across different inputs, and how they respond to perturbations. We find that the the \gwsnm ~machine learning model relies on specific input features to distinguish between glitch, NS and BBH sources.

In particular we show that the localization area of the 2D sky maps and the computed coherence versus incoherence Bayes factors are used as strong predictors for distinguishing between real (NS, BBH) events and glitches, and that the estimated distance to the source can further be used to discriminate between BBH events and mergers involving neutron stars. Contrary to expectations, the models do not learn to use the signal versus noise Bayes factors as a discriminant between real and glitch sources, but instead focus on coherence information inferred from \texttt{BAYESTAR} produced sky maps and metadata. The BBH-vs-all model is more reliant on the sky map localization shape and size to do these classifications as compared to the NS-vs-all model weighting the Log BCI and detector network inputs more heavily. The impact of the detector network shows a learned bias for associating events involving the Virgo detector more with real astrophysical events as compared to glitches, pointing to the need for updating our training set with more examples of Virgo glitches from O3.

After making predictions for the candidate events from O3, our perturbation results and insights into what the model is learning are also used to understand the misclassified events. When compared to the correctly classified events, the misclassified events are shown to have
distinct sky area, coherence factor, and distance values that influence the predictions. The observed trends are consistent with our expectations based on how \gwsnm ~leverages the sky area, distance, and coherence inputs, while the detector network reveals a subtle difference that is likely introducing a bias for the Virgo detector that we will address when re-training the model with O3 data.

Insights gained from our explainability studies help us to understand how the model works and makes its predictions, find its biases and limitations, and provide avenues for further optimization. With this context, users of \gwsnm ~can make more informed decisions for electromagnetic follow-up of candidate gravitational-wave events in LVK observing runs.

\section*{Acknowledgments}

The authors wish to acknowledge and highlight the contributions of Miriam Cabero in curating and developing the initial data and scripts that were used to train \gwsnm, which builds upon the \gwsn ~framework. The authors also wish to thank Aaron Tohuvavohu, Audrey Durand, Flavie Lavoie-Cardinal, and Ren\'{e}e Hlo\v{z}eck for their insights and discussions during a workshop in Montr\'{e}al in June 2022 that helped guide this work.

The authors highlight support for this project from the Canadian Tri-Agency New Frontiers in Research Fund – Exploration program. N.R. and N.V. acknowledge funding support from the Trottier Space Institute at McGill. D.H. and J.M. acknowledge support from the National Sciences and Engineering Research Council of Canada (NSERC) Discovery Grant program and the Canada Research Chairs (CRC) program. D.H. acknowledges support from the Canadian Institute for Advanced Research (CIFAR). A.M. acknowledges support from the NSF (1640818, AST-1815034). A.M. and J.M. also acknowledge support from IUSSTF (JC-001/2017). N.V. acknowledges funding from the NSERC Canada Graduate Scholarship - Doctoral (CGS-D), and the Murata Family Fellowship. This material is based upon work supported by NSF's LIGO Laboratory which is a major facility fully funded by the National Science Foundation.


\appendix

\section{Perturbation effects on accuracy}\label{sec:appA}
\begin{figure*}
\includegraphics[width=\textwidth]{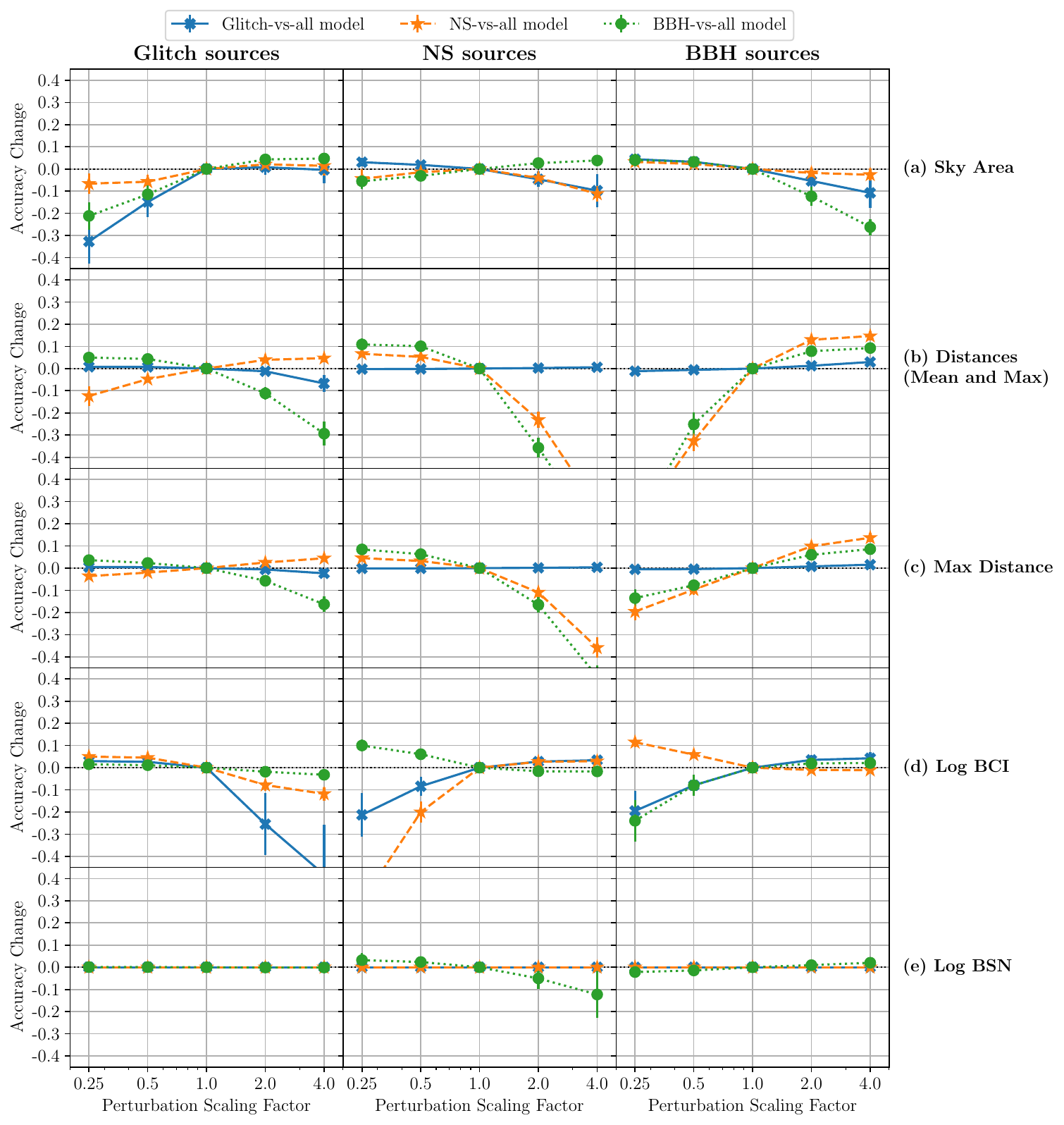}
\caption{\textit{Accuracy} change plots for the scaled input perturbations, providing a complementary view of the results shown in Figure~\ref{fig:fig4_prediction_score_change_scalar_all}. A negative accuracy score change of -0.2, for example, indicates that the model is misclassifying 20\% more events due to the perturbation.
\label{fig:fig8_accuracy_change_scalar_all}}
\end{figure*}

\begin{figure*}
     \centering
     \includegraphics[width=\textwidth]{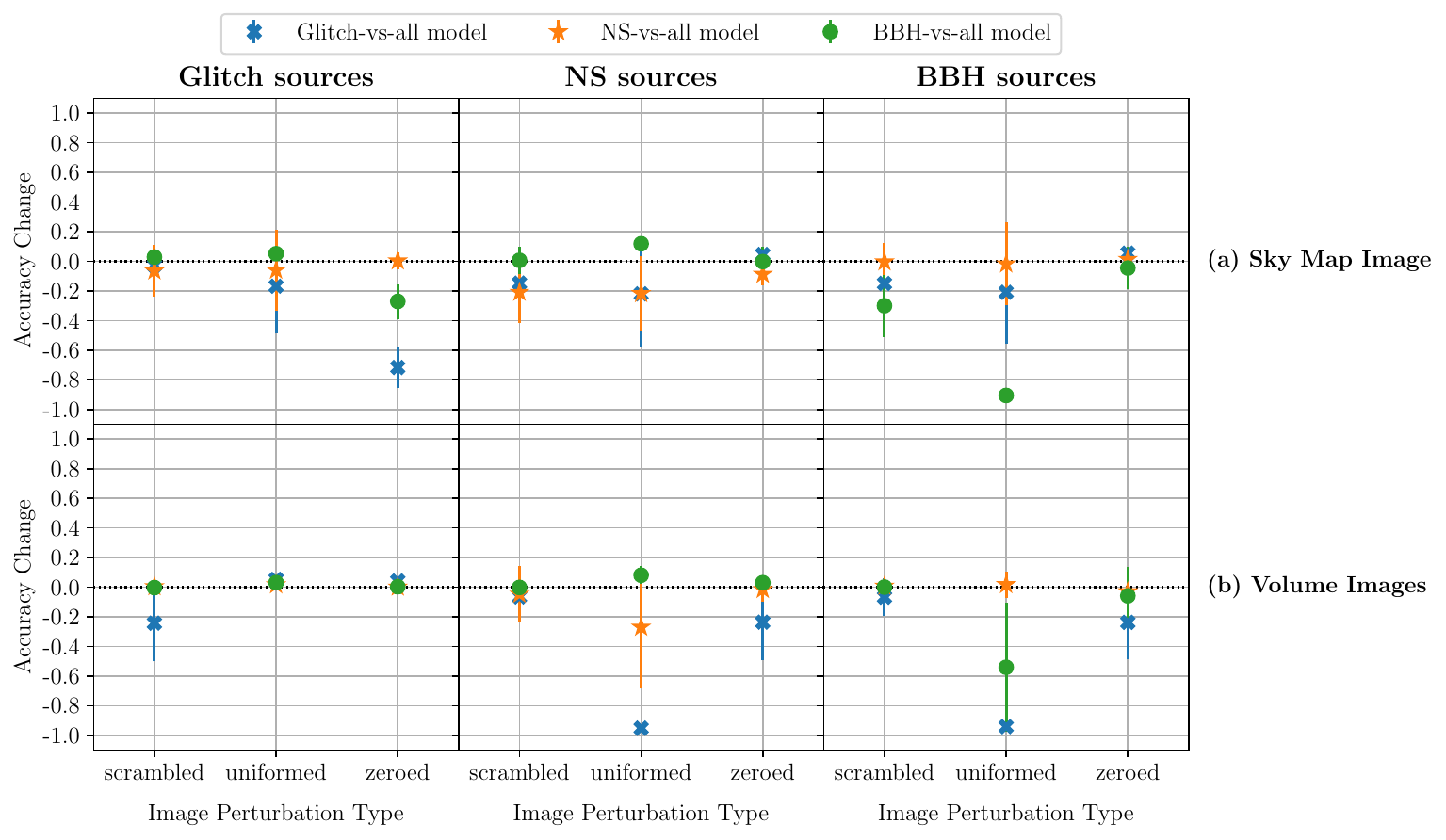}

     \bigskip
     \includegraphics[width=\textwidth]{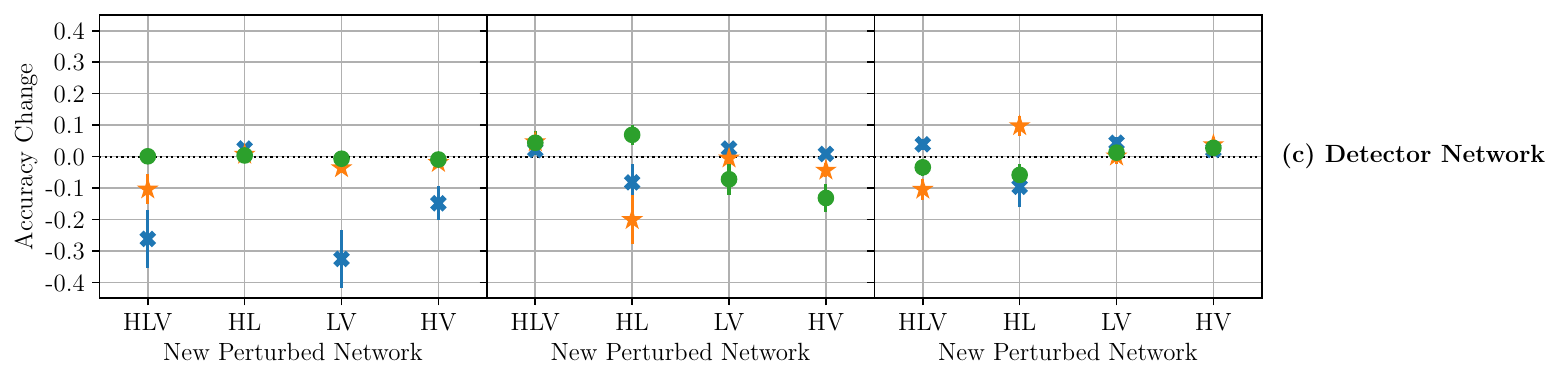}
    \caption{\textit{Accuracy} change plots for the un-scaled perturbations, providing a complementary view of the results shown in Figure~\ref{fig:fig5_prediction_score_change_nonscalar_all}}
    \label{fig:fig9_accuracy_change_nonscalar_all}
\end{figure*}

In this section we provide an alternate metric by which to study the effects of the input perturbations. This metric is based on how the percentage of events that are accurately classified in the testing set changes when the input is perturbed, so that now we are not looking at how much the predictions score changes, but in how many cases is that score change large enough to exceed the threshold value and classify the event differently. The threshold values are chosen for each model as in \cite{Abbott2022}, i.e., the value between 0 and 1 for which the false alarm rate and false positive rate is equal for that model. The accuracy with the perturbed input is then compared to the accuracy with the unperturbed original input to measure the accuracy change. This is repeated for the 20 model iterations and used to calculate the mean accuracy change value and standard deviation. The results for each input perturbation type for the scalar perturbations is shown in Figure~\ref{fig:fig8_accuracy_change_scalar_all}, and for the image-based and detector network perturbations shown in Figure~\ref{fig:fig9_accuracy_change_nonscalar_all}.

\section{Extended look at Log BSN}\label{sec:appB}

\begin{figure*}
\includegraphics[width=\textwidth]{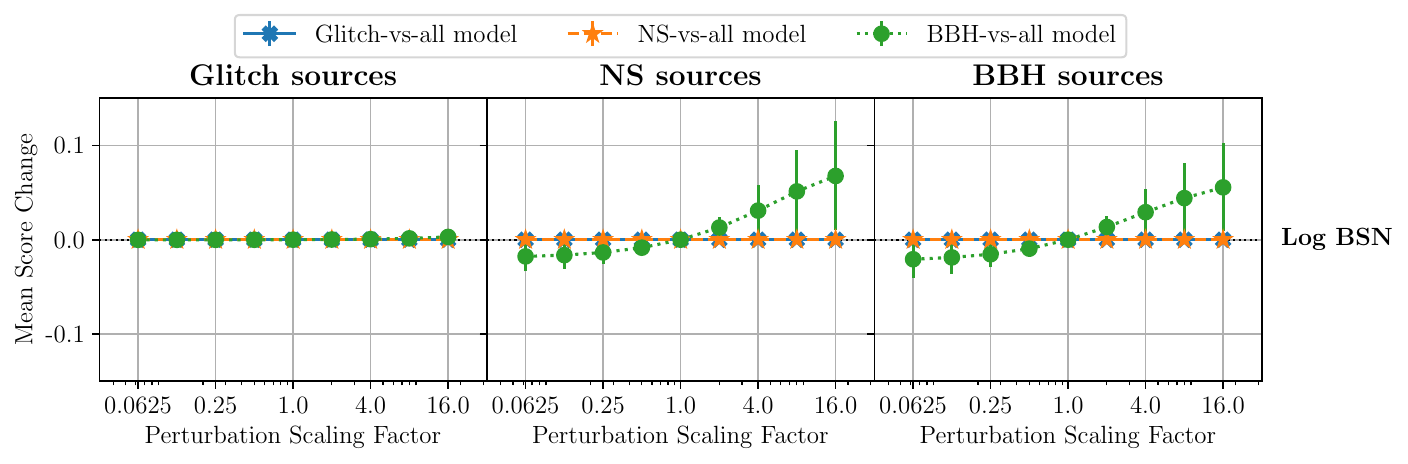}
\caption{Extended perturbation score change plot for Log BSN. The slight dependence of the BBH-vs-all classifier for the real sources is visible, but there is no change otherwise, confirming that the \gwsnm ~predictions are not sensitive to the Log BSN input value.
\label{fig:fig10_prediction_score_change_logBSN_extended}}
\end{figure*}

To see if any change in score would arise for more extreme perturbations to the Log BSN factor, we show in Figure~\ref{fig:fig10_prediction_score_change_logBSN_extended} the results for perturbing the Log BSN values by factors of up to 16 times larger and smaller. Even then the results are the same as those seen in Figure~\ref{fig:fig4_prediction_score_change_scalar_all}, with the slight dependence of the BBH-vs-all classifier for the real sources visible, but no changes otherwise, confirming that the \gwsnm ~predictions are not sensitive to the Log BSN input value.

\begin{figure}
\centering
\includegraphics{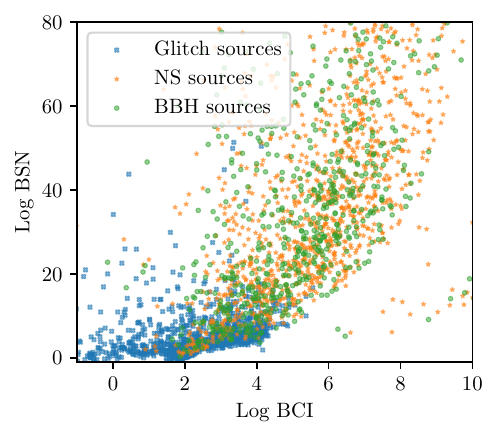}
\caption{Scatter plot of the Log BSN versus the Log BCI value for all 4267 events in our data set. The scatter shows a clear positive correlation between the two input values for all source types. This helps explain why the trends in Figure~\ref{fig:fig5_prediction_score_change_nonscalar_all} seem to show a model dependence on the Log BSN values, even though the perturbation results show that it has no effect: the predictions depend on the Log BCI value, and thus correlate with the Log BSN value.
\label{fig:fig11_BSN_BCI_correlation_scatter}}
\end{figure}

The apparent trend for the Log BSN factor seen in Figure~\ref{fig:fig3_original_prediction_score_all}(e) is instead likely due to the fact that the Log BSN value is correlated with the Log BCI value and it is the latter which the models are dependent on. We show this correlation in Figure~\ref{fig:fig11_BSN_BCI_correlation_scatter}. When \texttt{BAYESTAR} is modeling the detector network response to a signal, it has more support for the signal to be coherent across detectors (larger Log BCI value) when it is louder and easier to distinguish from the background noise (larger Log BSN value).

\section{O3 misclassifications}\label{sec:appC}
The list of events in O3 that are misclassified by \gwsnm, along with select parameter values of these events that are input to \gwsnm ~and found to be used by the model in making its classification prediction, are shown in Table ~\ref{table:O3_misclass}. The data in the table (except for the event name and the detector network) are shown graphically in Figure~\ref{fig:fig7_O3_classifications_vs_input_all}.

\begin{table*}
\centering
\renewcommand{\arraystretch}{1.2}
\begin{tabular}{ c c c c c c c }
\hline
\hline
\textbf{Event ID} & \textbf{Detectors} & \textbf{90\% Sky Area} & \textbf{Mean Distance} & \textbf{Log BCI} & \textbf{GWTC-3} & \textbf{GWSkyNet-Multi} \\
\textbf{} & \textbf{} & $\mathbf{[deg^2]}$ & \textbf{[Mpc]} & \textbf{} & \textbf{Classification} & \textbf{Classification} \\
\hline
\hline
S190405ar & HLV & 2677 & 268 & 1.9 & Glitch & NS \\
\hline
S190425z & LV & 10183 & 155 & 0.8 & BNS & Glitch \\
\hline
S190426c & HLV & 1262 & 375 & 8.1 & Glitch & NS \\
\hline
S190503bf & HLV & 448 & 421 & 6.5 & BBH & NS \\
\hline
S190630ag & LV & 8493 & 1059 & 1.1 & BBH & Glitch \\
\hline
S190816i & LV & 1467 & 261 & 4.2 & Glitch & NS \\
\hline
S190923y & HL & 2107 & 438 & 5.8 & Glitch & NS \\
\hline
S190924h & HLV & 515 & 514 & 7.6 & BBH & NS \\
\hline
S191213g & HLV & 1393 & 195 & 3.9 & Glitch & NS \\
\hline
S191216ap & HV & 300 & 324 & 4.8 & BBH & NS \\
\hline
S191220af & LV & 5238 & 166 & 3.0 & Glitch & NS \\
\hline
S191225aq & LV & 1154 & 24 & 6.4 & Glitch & NS \\
\hline
S200106au & HL & 207 & 48 & 9.8 & Glitch & NS \\
\hline
S200112r & LV & 6199 & 1136 & 0.0 & BBH & Glitch \\
\hline
S200302c & HV & 6705 & 1737 & 0.8 & BBH & Glitch \\
\hline
\hline
\end{tabular}
\caption{Details of the 15 events misclassified by \gwsnm ~in the LVK third observing run (O3). The columns are: the candidate event ID on GraceDB, detectors that were observing at the time of the event, 90\% credible sky localization area of the \texttt{BAYESTAR} generated sky map, estimated mean posterior distance, log Bayes factor of coherence vs incoherence model, GWTC-3 classification of the event, and \gwsnm ~predicted classification. The GWTC-3 classifications are obtained by considering only events with $p_{astro} > 0.5$ as real events (with the rest classified as glitches). The type of binary for the real events is determined from the best-fit component masses, assuming a maximum neutron star mass of $\mathrm{3~M_{\odot}}$.}
\label{table:O3_misclass}
\renewcommand{\arraystretch}{1}
\end{table*}

\bibliography{references}{}
\bibliographystyle{aasjournal}

\end{document}